\documentclass[prb,aps,twocolumn,showpacs,superscriptaddress]{revtex4}
\usepackage{epsfig}
\usepackage{amsmath}
\usepackage{amsfonts}
\usepackage{mathptmx}
\usepackage{eucal}

\DeclareMathOperator{\Tr}{\mathop{\mathrm{Tr}}}
\DeclareMathOperator{\sgn}{\mathop{\mathrm{sgn}}}
\DeclareMathOperator{\re}{\mathop{\mathrm{Re}}}
\DeclareMathOperator{\im}{\mathop{\mathrm{Im}}}
\DeclareMathOperator{\arctanh}{arctanh}
\DeclareMathOperator{\const}{\mathop{\mathrm{const}}}
\newcommand{\Eq}[1]{Eq.~(\ref{#1})}
\newcommand{\Eqs}[1]{Eqs.~(\ref{#1})}

\begin{document}

\title{Nonequilibrium effects in tunnel Josephson junctions}

\author{E.V.~Bezuglyi}
\affiliation{Institute for Low Temperature Physics and Engineering,
Kharkov 61103, Ukraine}
\affiliation{Chalmers University of Technology, S-41296
G\"oteborg, Sweden}

\author{A.S.~Vasenko}
\affiliation{Chalmers University of Technology, S-41296
G\"oteborg, Sweden}
\affiliation{Department of Physics, Moscow State University,
Moscow 119992, Russia}

\author{V.S.~Shumeiko}
\affiliation{Chalmers University of Technology, S-41296
G\"oteborg, Sweden}

\author{G.~Wendin}
\affiliation{Chalmers University of Technology, S-41296
G\"oteborg, Sweden}

\date{\today}

\begin{abstract}
We study nonequilibrium effects in current transport through
voltage biased tunnel junction with long diffusive superconducting
leads at low applied voltage, $eV \ll 2\Delta$, and finite
temperatures. Due to a small value of the Josephson frequency, the
quasiparticle spectrum adiabatically follows the time evolution of
the superconducting phase difference, which results in the
formation of oscillating bound states in the vicinity of the
tunnel junction (Andreev band). The quasiparticles trapped by the
Andreev band generate higher even harmonics of the Josephson ac
current, and also, in the presence of inelastic scattering, a
nonequilibrium dc current, which may considerably exceed the dc
quasiparticle current given by the tunnel model. The distribution
of travelling quasiparticles also deviates from the equilibrium
due to the spectrum oscillations, which results in an additional
contribution to the dc current, proportional to $\sqrt{V}$.
\end{abstract}


\maketitle

\section{Introduction}

Tunnel current transport in superconducting junctions is a
classical topic of interest and research.\cite{review} Theoretical
description of the superconducting tunneling given by the tunnel
model\cite{Wert,LO} is based on the assumption about equilibrium
in the junction electrodes. The resulting equations for the tunnel
current are expressed through the nonper\-tur\-bed density of
states (DOS) and equilibrium distribution function. Such an
approach includes only single-particle tunnelling processes, and
it is sufficient for describing current-voltage characteristics at
large applied voltage $eV > 2\Delta$, and within the subgap
voltage region $eV < 2\Delta$ at nonzero temperature. At very
small temperature, multiparticle tunneling processes must be taken
into account.\cite{MPT,Bratus95} However, these processes are
exponentially weak at small voltage, and if the temperature is not
particularly small, $eV\ll T < 2\Delta$, only tunneling of
thermally excited quasiparticles plays significant role. It is
clear, however, that quasiparticle tunnelling generates
nonequi\-li\-b\-ri\-um distribution in the electrodes, and that
the effect is enhanced in diffusive junctions because of the
scattering by impurities. Traditionally, this effect is considered
to be negligibly small since it is of a higher order in the
junction transparency.

\begin{figure}[tb]
\epsfxsize=8.0cm\epsffile{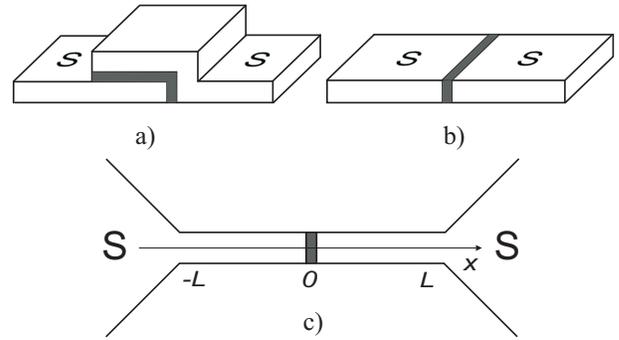} \vspace{-2mm}
\caption{Possible experimental realizations of tunnel junctions
(a,b) and the theoretical model (c): the tunnel barrier attached
to bulk superconducting electrodes by long ($L \gg \xi_0$)
diffusive superconducting leads.} \label{model} \vspace{-4mm}
\end{figure}

How small is the nonequilibrium effect actually? In voltage biased
superconducting junctions, the phase of the order parameter has
different time dependencies in different electrodes, and the
interference of the order parameters induced by the tunnelling
leads to the time oscillations of the DOS. The character of these
oscillations can be qualitatively understood from comparison with
the well studied case of ballistic tunnel
junctions.\cite{Fur,Averin95,Bratus97} In the latter, the Andreev
levels are formed in the vicinity of the tunnel barrier; their
energies lie within the energy gap in the bulk electrodes and
oscillate in time following evolution of the superconducting phase
difference. Similarly, one may expect that in diffusive junctions
the time oscillation of the DOS will have the form of a
``breathing'' potential well localized in the vicinity of the
tunnel barrier (``Andreev band''), which periodically, with the
Josephson period, traps and releases quasiparticles. Thus the
quasiparticles spend larger time within the junction area compared
to the travelling time, which should generate strongly
nonequilibrium quasiparticle distribution and, as a result, an
appreciable change of the tunnelling current. The effect should be
most pronounced for a planar junction geometry sketched in
Fig.~\ref{model}(a,b), when the tunnel junction is connected to
bulk electrodes via diffusive superconducting wires whose length
exceeds the size of the Andreev band. Otherwise, as in diffusive
point contacts, the proximity of large equilibrium reservoirs will
suppress the DOS oscillations and hence formation of the Andreev
band because of rapid spreading out of the current.

In this paper we demonstrate that the nonequilibrium quasiparticle
distribution generated by nonstationary process of formation of
the Andreev band during the tunneling considerably modifies the dc
tunnel current at small applied voltage. We show that the
nonequilibrium effect results in a time dependent amplitudes of
all the three current components given by the tunnel
model,\cite{Wert,LO}
\begin{equation}\label{Itun}
I(t) = I_1(t)\sin\phi  + I_2(t) + I_3(t) \cos\phi, \;\;\; \phi= 2eVt
\end{equation}
(Josephson current, quasiparticle current, and the interference
current, respectively). All the amplitudes have nonharmonic time
dependence, and generally all of them contribute to the dc
current. However, at small voltage, the major contribution,
nonanalytic in junction transparency and voltage, comes from the
sine term (Josephson current). More precisely, this additional dc
current results from a non-adiabatic component of the distribution
function associated either with inelastic relaxation of
quasiparticles, or with the quasiparticle diffusion away from the
tunnel barrier. The additional dc current may considerably exceed
the magnitude given by the tunnel model, and moreover, leads to a
{\em non-monotonous} net dc current.

Such effect has close qualitative similarity to the
nonequi\-lib\-rium effects in transparent weak links
[superconductor-nor\-mal metal-superconductor (SNS) junctions and
superconducting bridges] studied earlier.\cite{bridges,longSNS}
There are however the differences: In superconducting bridges, the
potential well appears due to suppression of the order parameter
by the current concentration (depairing effect),\cite{bridges}
while in the tunnel junctions this effect does not play any
essential role. In long SNS junctions, the potential well is
pre-prepared by the proximity effect, and it exists in the absence
of the transport current; as a result, the oscillations develop at
small energy,\cite{longSNS} while in the tunnel junctions the
oscillations occur at large energies close to the gap edge in the
electrodes.

Organization of the paper is the following. In Sec.\ II we
formulate the theoretical model and microscopic equations
describing current transport through the tunnel junction with
diffusive leads. In Sec.\ III we introduce an adiabatic approach
for solving the time-de\-pen\-dent problem in the limit of low
applied voltage. This approach is applied to calculation of the
spectral characteristics of the junction (Sec.\ IV) and the
quasiparticle distribution function (Sec.\ V). In Sec.\ VI we
calculate the ac and dc components of the net Josephson current,
and then discuss and summarize the results in Sec.\ VII and VIII.

\section{Model and Basic equations}

The model of tunnel junction we are going to study is depicted in
Fig.~\ref{model}(c) and consists of a tunnel barrier with the
transparency $\Gamma$ attached to bulk super\-conducting
electrodes via two superconducting leads ($-L < x <0$ and
$0<x<L$). We will consider diffusive limit, in which the elastic
scattering length $\ell$ is much smaller than the coherence length
$\xi_0 = \sqrt{\mathcal{D}/2\Delta}$, where $\mathcal D$ is the
diffusion coefficient (we assume $\hbar = k_B = 1$). The length
$L$ of the leads is assumed to be much larger than $\xi_0$ (long
junction), and their width will be supposed to be much smaller
than the Josephson penetration depth which implies homogeneity of
the current along the junction. Similar model was considered in
Refs.~\onlinecite{Kupr}  and \onlinecite{BBG} in study of the dc
Josephson effect in tunnel structures.

Under these conditions, the microscopic calculation of the
electric current $I(t)$ requires solution of the one-dimensional
diffusive equations of nonequilibrium
super\-conductivity\cite{LOnoneq} (see also a review\cite{Belzig})
for the $4 \times 4$ matrix two-time Keldysh-Green's function
$\check{G}(x,t_1,t_2)$ in the leads,
\begin{align}
\left[\check{H}, \check{G}\right] &= i \mathcal{D}
\partial_x\check{J}, \;\; \check{J} = \check{G}\circ\partial_x
\check{G}, \;\; \check{G}^2 = \delta(t_1 - t_2),\label{GK}
\\ \nonumber
\check{H} &= \bigl[i\sigma_z \partial_{t_1} - e\varphi +
\hat{\Delta}(t_1)\bigr]\delta(t_1-t_2),
\end{align}
where $\hat{\Delta} = e^{i\sigma_z\phi}i\sigma_y\Delta$, $\varphi$
is the electric potential, $\Delta$ and $\phi$ are the modulus and
the phase of the order parameter, respectively, $\sigma_i$ are the
Pauli matrices, and
\begin{equation}
\check{G} = \begin{pmatrix} \hat{g}^R & \hat{G}^K \\
0 & \hat{g}^A
\end{pmatrix}, \quad \hat{G}^K = \hat{g}^R \circ\hat{f} - \hat{f}\circ
\hat{g}^A. \label{defG}
\end{equation}
In \Eq{defG}, $\hat{g}^{R,A}$ are the $2 \times 2$ Nambu matrix
retarded and advanced Green's functions, and $\hat{f} = f +
\sigma_z f_-$ is the matrix distribution function (we use `check'
for $4 \times 4$ and `hat' for $2 \times 2$ matrices). The
multiplication procedure in \Eqs{GK}-\eqref{defG} is defined as
the time convolution,
\begin{align}
(A\circ B)(t_1, t_2) = \int^{+\infty}_{-\infty} A(t_1, t)B(t, t_2)
dt.\nonumber
\end{align}
In \Eqs{GK}, we neglect the inelastic collision term, which will
be taken into account later, in Sec.~VI.

The boundary conditions for the function $\check{G}$ and the
matrix current $\check{J}$ at the left ($x=-0$) and the right
($x=+0$) sides of the tunnel junction are given by
relation\cite{KL}
\begin{equation}
\check{J}_{-0} = \check{J}_{+0} = (2g_N R)^{-1}
\bigl[\check{G}_{-0}, \check{G}_{+0}\bigr],\label{KL}
\end{equation}
where $R$ is the junction resistance and $g_N$ is the normal
conductivity of the leads per unit length. The electric current is
related to the Keldysh component of the matrix current $\check{J}$
as
\begin{equation}
I(t)=({\pi  g_N}/{4e}) \Tr \sigma_z \hat{J}^K(x,t,t),
\nonumber
\end{equation}
and thus it can be expressed through the boundary value
$\check{J}_0$,
\begin{equation}
I(t)=\frac{\pi }{8eR} \Tr \sigma_z
\bigl[\check{G}_{-0},{\check{G}_{+0}}\bigr]^K(t,t). \label{Curr2}
\end{equation}

Equations (\ref{GK}) can be decomposed into the diffusion equations
for the Green's functions,
\begin{equation}
\bigl[\hat{H}, \hat{g}\bigr] = i \mathcal{D}\partial_x \hat{J}, \;\;
\hat{J} = \hat{g}\circ\partial_x \hat{g}, \;\; \hat{g}^2 = \delta(t_1
- t_2),\label{Usadel}
\end{equation}
and the equation for the Keldysh component $\hat{G}^K$,
\begin{equation}
\bigl[\hat{H}, \hat{G}^K\bigr] = i \mathcal{D}\partial_x
\hat{J}^K, \; \hat{J}^K = \hat{g}^R \circ\partial_x\hat{G}^K +
\hat{G}^K \circ\partial_x \hat{g}^A. \label{KineticG}
\end{equation}
The boundary conditions for the functions $\hat{g}$ and
$\hat{G}^K$ at the tunnel barrier follow from \Eq{KL},
\begin{subequations} \label{Boundary2}
\begin{align}
\hat{J}_{0}  &= (W / \xi_0)\bigl[\hat{g}_{-0}, {\hat{g}}_{+0}\bigr],
\label{BoundaryG}
\\
\hat{J}_{0}^K  &= (W / \xi_0)\bigl[\check{G}_{-0},
{\check{G}}_{+0}\bigr]^K.
 \label{BoundaryGK}
\end{align}
\end{subequations}

In \Eqs{Boundary2}, the transparency parameter $W$ is defined as
\begin{equation} \label{W}
W = {R(\xi_0)}/{2R} = ({3\xi_0}/{4\ell})\Gamma \gg \Gamma, \nonumber
\end{equation}
where $R(\xi_0) = \xi_0/g_N$ is the normal resistance of the lead
per length $\xi_0$. It has been shown in Refs.~\onlinecite{Kupr}
and \onlinecite{BBG} that this quantity rather than the barrier
transparency $\Gamma$ plays the role of a true transparency
parameter in the theory of diffusive tunnel junctions (see also
discussion in Sec.\ VII). In this paper, we will consider the
limit $W \ll 1$, which corresponds to the conventional tunneling
concept. In this case, according to \Eqs{Boundary2}, the gradients
of all functions are small. Within the tunnel model, which assumes
$W$ to be the smallest parameter in the theory, these gradients
are neglected, and the functions $\hat{g}$ and $\hat{f}$ are
assumed to be local-equi\-librium within the leads. In our
consideration, we will lift this assumption and suppose the
local-equi\-librium form of these functions only within the bulk
electrodes (reservoirs). Attributing the reference point for the
phase, $\phi=0$, to the left electrode, $x=-L$, these functions in
the right electrode, $x=L$, are given by relations,
\begin{subequations} \label{Equil}
\begin{align}
&\hat{g}(E,t) =  \sigma_z u_S(E+\sigma_z eV) +
ie^{i\sigma_z\phi(t)}\sigma_y v_S(E) , \label{gS}
\\ \nonumber
&(u_S,v_S)^{R,A} = \frac{(E,\Delta)}{(E_{R,A}^2-\Delta^2 )^{1/2}},
\;\; E_{R,A}=E\pm i0,
\\
&\hat{f}(E) = n_{\textit{eq}}(E+\sigma_z eV), \quad
n_{\textit{eq}}(E)=\tanh (E/2T), \label{f0}
\end{align}
\end{subequations}
written in terms of the mixed Wigner representation $A(E,t)$ of
the two-time functions,
\begin{align} \label{Wigner}
A(t_1, t_2) &= \int_{-\infty}^{+\infty} \frac{dE}{2 \pi }\, e^{-iE
(t_1 - t_2)} A(E, t).
\end{align}
In \Eq{Wigner}, the variable $E$ has the meaning of the
quasiparticle energy, and $t = (t_1 + t_2)/2$ is a real time.
Similar equations, with $\phi=0$ and $V=0$, apply to the left
electrode, $x=-L$.

Because of the small value of the tunneling parameter $W$ one can
neglect the variation of the electric potential and the
superconducting phase along the leads, and assume the voltage $V$
and the phase difference $\phi = 2eV t$ between the reservoirs to
be directly applied to the tunnel barrier. Following this
argument, one can also neglect the variation of the charge
imbalance function $f_-$ proportional to a small electric field
($\sim eVW$) penetrating the superconducting leads. Furthermore,
the small value of the superfluid momentum in the superconducting
leads,\cite{Kupr,BBG} $p_s \sim W/\xi_0$, enables us to neglect a
small effect of the suppression of the energy gap,\cite{LO1}
[$\sim (p_s\xi_0)^{4/3} \sim W^{4/3}$], by the superfluid
momentum. In such approximation, the coefficients in the equation
\eqref{GK} within the left lead, $x<0$, are time-independent
functions, similar to the value of $\check{G}$ at the left
electrode. At $x>0$, using the gauge transformation,
$\widetilde{\check{G}}(x>0,t_1,t_2) =
S^\dagger(t_1)\check{G}(x>0,t_1,t_2) S(t_2)$,  of the function
$\check{G}$ with a unitary operator $S(t)
=\exp[i\sigma_z\phi(t)/2]$,\cite{Artemenko} we exclude the
time-dependent phase and the electric potential from the equations
for the function $\widetilde{\check{G}}$ and the boundary
conditions \eqref{Equil} at $x=L$, which then become similar to
the equations for $\check{G}(x)$ at $x<0$ and the boundary
conditions at $x=-L$. This results in the symmetry relation
$\widetilde{\check{G}}(x) = \check{G}(-x)$, which allows us to
replace the function $\check{G}_{+0}$ in the boundary condition
\eqref{KL} and in the expression \eqref{Curr2} for the electric
current by the inverse gauge transformation of the function
$\check{G}_0 \equiv \check{G}_{-0}$,
\begin{equation}
\check{G}_{+0} \to \overline{\check{G}}_0 \equiv   S(t_1)
\widetilde{\check{G}}_{+0} S^\dagger(t_2) = S(t_1) \check{G}_0
S^\dagger(t_2). \label{DefBar}
\end{equation}
As the result, the problem is reduced to the solution of a static
equation for the function $\check{G}(x)$ within the left lead with
the time-dependent boundary condition \eqref{KL} at the tunnel
barrier. Similar approach is used in the theory of ballistic point
contacts, where the Josephson coupling is described by an
effective time-dependent matching condition for the
gauge-trans\-formed Bogolyubov-de Gennes equations in the leads.

In a general nonstatic case, the function $\check{G}(t_1,t_2)$
consists of a set of harmonics $\check{G}(E_n,E_m)$, $E_n =
E+neV$, which are coupled to each other through a complicated set
of recursive equations following from \Eqs{GK} (see discussion in
Ref.~\onlinecite{Brinkman}). The problem essentially simplifies if
the distance $eV$ between the harmonics is much smaller than the
smallest scale $\delta E$ of variations in the quasiparticle
spectrum,
\begin{equation} \label{Cond}
eV \ll \delta E.
\end{equation}
The magnitude of $\delta E$ will be indicated below, see
\Eq{depth}.

\section{Low voltage limit}

In the limit of low applied voltages, the evolution of the
quasiparticle spectrum and distribution function can be described
within the adiabatic approximation using expansion over small
Josephson frequency. In the static case, $eV \to 0$, the function
$\check{G}$ depends only on the time difference $t_1-t_2$, and the
Wigner transformation \eqref{Wigner} reduces \Eqs{Usadel} to the
standard Usadel equations,\cite{Usadel}
\begin{subequations} \label{UsadelSt}
\begin{align}
E v - \Delta u &= (i\mathcal{D}/2)\partial_x\left(u \partial_x v - v
\partial_x u\right), \label{Usadel1}
\\
u^2 - v^2 &= 1, \label{Usadel2}
\end{align}
\end{subequations}
for the scalar components of the Green's function
\begin{equation} \label{static}
\hat{g}(x,E) = \sigma_z u + i\sigma_y v.
\end{equation}
The functions $u$ and $v$ determine the spectral characteristics
of the system; in particular, the quantity $N(x,E)= (u^R-u^A)/2$
is the DOS normalized over its value $N_F$ in the normal state. In
what follows, we will express the advanced Green's functions
through the retarded ones, $(u,v)^{A} = - (u,v)^{R\ast}$, using
the general relation $\hat{g}^A = -\sigma_z
\hat{g}^{R\dagger}\sigma_z$, and omit the superscript $R$,
assuming all Green's functions to be retarded.

At small applied voltages, we proceed to the Wigner representation
\eqref{Wigner} of the two-time functions and expand the time
convolutions in \Eqs{Usadel}-\eqref{Boundary2} to first order in
$eV$,
\begin{equation} \nonumber
(A \circ B)(E, t) \approx A B + {(i /2)}\langle A, B\rangle,
\end{equation}
where $\langle A, B \rangle = \partial_E A \,\partial_t B -
\partial_t A \, \partial_E B$ denotes the Poisson brackets in the
energy-time space. Within such approximation, the Green's function
$\hat{g}(x,E,t)$ holds the matrix structure \eqref{static}, and
the gauge-transformed functions $\overline{\hat{g}}_0$ and
$\overline{f}_0$ [see \Eq{DefBar}], which enter the boundary
conditions \eqref{Boundary2}, read
\begin{subequations} \label{overline}
\begin{align}
\overline{\hat{g}}_0 &=\sigma_z u_0(E+\sigma_z eV,t) +
ie^{2i\sigma_z eV t} \sigma_y v_0(E,t), \label{overlineg}
\\
\overline{f}_0&=f_0(E+\sigma_z eV,t). \label{overlinef}
\end{align}
\end{subequations}

The expression for the electric current obtained from \Eqs{Curr2} and
\eqref{overline} consists of the three terms,
\begin{subequations}\label{currents}
\begin{align}
I(t) &= I_1\sin\phi  + I_2 + I_3 \cos\phi, \label{I}
\\
I_1(t)\sin\phi &= \frac{1}{2eR}\int_{-\infty}^\infty dE\, I_s(E,t)
f_0(E,t) , \label{I1}
\\
I_2(t) &= \frac{V}{R}\int_0^\infty dE\, N_0^2(E,t) \,\partial_E
f_0(E,t), \label{I2}
\\
I_3(t) &= \frac{V}{R}\int_0^\infty dE\, M_0^2(E,t) \,\partial_E
f_0(E,t),\label{I3}
\end{align}
\end{subequations}
where $f_0(E,t)$ is the boundary value of the distribution
function $f(x,E,t)$, and $I_s= - \im v_0^2 \sin \phi$, $N_0 \equiv
N(x=0,E)= \re u_0$, and $M_0=\re v_0$ are the spectral densities
of the different current components. Such structure of the current
is similar to the result of the tunnel model:\cite{LO} indeed,
when the current spectral densities and distribution function
approach nonperturbed equilibrium form, the first term in \Eq{I}
describes the ac Josephson current, the term $I_2$ is the
dissipative quasiparticle current which approaches the value $V/R$
in the normal state, and $I_3$ is the interference current.
However, \Eqs{currents} are more general, they include the
Josephson oscillations of the spectral characteristics together
with a nonequilibrium form of the distribution function.

\begin{figure}[t]
\epsfxsize=8.5cm\epsffile{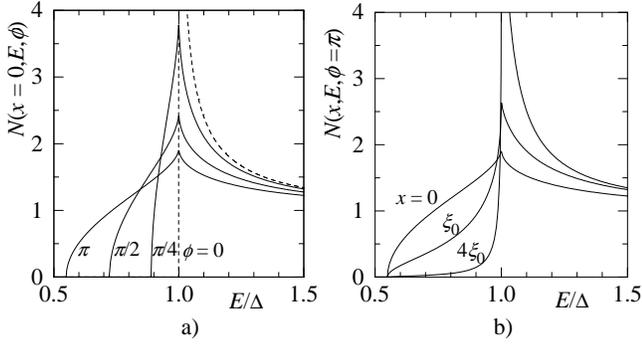} \vspace{-2mm}
\caption{Energy dependencies of the DOS for the transparency
parameter $W=0.05$: (a) at the tunnel barrier for different values
of the superconducting phase; (b) at $\phi=\pi$ for different
distances from the barrier. }
\label{dosex} \vspace{-3mm}
\end{figure}

\section{Junction spectrum}

To first order in $eV$, the spectral functions $u(x,E,t)$ and
$v(x,E,t)$ obey static Usadel equations \eqref{UsadelSt} with
time-de\-pen\-dent boundary condition following from
\Eqs{BoundaryG} and \eqref{overlineg},
\begin{equation}
\xi_0\left(u \partial_x v - v \partial_x u\right)_0 =
-4W(uv)_0\sin^2(\phi/2). \label{Boundary1}
\end{equation}
This boundary condition was found in Refs.~\onlinecite{Kupr} and
\onlinecite{BBG} for a similar structure with the time-independent
phase difference. Thus, at low enough voltages, the spectral
characteristics of the junction adiabatically follow time
variations of $\phi(t)$.

Equations \eqref{UsadelSt} and \eqref{Boundary1} can be supplemented
by helpful identities following from the unity components of the
matrix equations \eqref{Usadel} and \eqref{BoundaryG},
\begin{subequations}
\begin{align}
&\partial_t u = (i\mathcal{D}/2)\partial_x \bigl(\left\langle u,
\partial_x u\right\rangle - \left\langle v,
\partial_x v\right\rangle\bigr), \label{Equal1}
\\
&\xi_0 \bigl(\left\langle u, \partial_x u\right\rangle -
\left\langle v,
\partial_x v\right\rangle\bigr)_0 = -W  \langle v^2_0,
\cos\phi \rangle. \label{Equal2}
\end{align}
\end{subequations}

In order to satisfy the condition \eqref{Usadel2}, we introduce
usual parametrization $u = \cosh\theta$, $v =\sinh\theta$.
Furthermore, we will neglect a small effect of suppression of the
order parameter near the barrier, assuming $\Delta$ to be
homogeneous [see comment to \Eq{depth}]. Then the equation and the
boundary condition for the spectral angle $\theta$, following from
\Eqs{Usadel1} and \eqref{Boundary1}, take the form in a
dimensionless variable $z = x/\xi_0$,
\begin{align}
&\partial_z \theta = 2 k\sinh\frac{\theta - \theta_S}{2}, \quad
k(E) = [i\sinh\theta_S(E)]^{-1/2}, \label{EqTheta}
\\
&\left(\partial_z \theta\right)_0 = -2\gamma(t)\sinh 2\theta_0,
\quad \gamma(t)=W \sin^2[\phi(t)/2]. \label{BoundaryTheta1}
\end{align}

The quantity $\gamma(t)$ in \Eq{BoundaryTheta1} has the meaning of
a time-dependent depairing factor related to the discontinuity of
the superconducting phase $\phi(t)$ at the tunnel barrier. When
the phase approaches multiple of $2\pi$, this factor turns to
zero, and the spectral angle becomes homogeneous and equal to its
bulk value $\theta_S(E) = \arctanh ({\Delta / E})$. At arbitrary
$\phi$, equations \eqref{EqTheta} and \eqref{BoundaryTheta1}
describe deviation of the spectral angle from $\theta_S$ at the
distance $x \sim \xi_0$ from the barrier; thus, in a long
junction, $L \gg \xi_0$, we can apply the solution for a
semi-infinite lead,\cite{BV}
\begin{equation}
\tanh\bigl[(\theta - \theta_S)/{4}\bigr] = \tanh\bigl[(\theta_0 -
\theta_S)/{4}\bigr]\exp(k z). \label{SolTheta}
\end{equation}
\begin{figure}[t]
\epsfxsize=8.0cm\epsffile{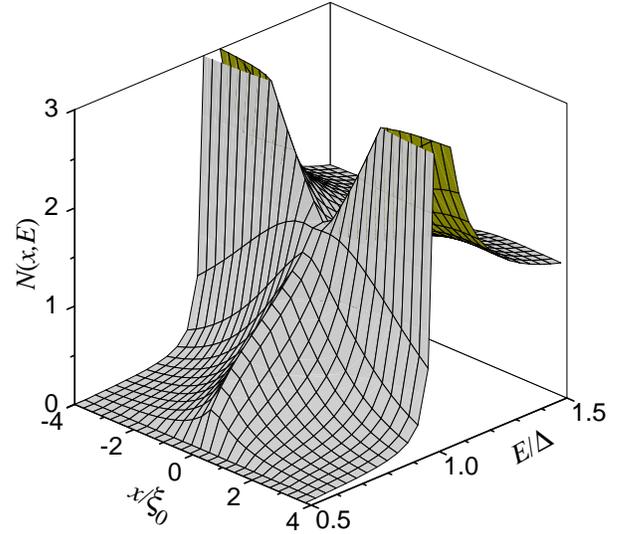} \vspace{-2mm}
\caption{2D profile of the DOS in the vicinity of the tunnel
barrier at $W=0.05$, at the moments of maximum depth of the
Andreev band ($\phi = \pi$). }
\label{dos2d} \vspace{-3mm}
\end{figure}

The equation for the boundary value $\theta_0(E,t)$ of the
spectral angle follows from \Eqs{EqTheta}-\eqref{SolTheta},
\begin{equation}
\sinh\frac{\theta_S(E) - \theta_0(E,t)}{2} =
\frac{\gamma(t)}{k(E)}\sinh 2\theta_0(E,t). \label{EqTheta0}
\end{equation}

\begin{figure}[tb]
\epsfxsize=8.5cm\epsffile{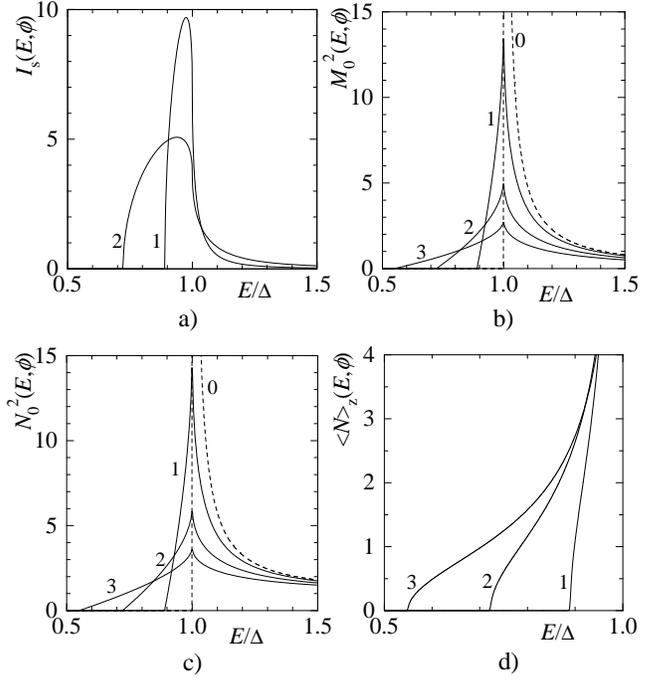} \vspace{-2mm}
\caption{Spectral characteristic of the junction at $W=0.05$, for
different values of the su\-per\-con\-duc\-ting phase: $\phi=0$
(curves indicated by 0), $\pi/4$ (1), $\pi/2$ (2), and $\pi$ (3).
(a-c) -- spectral densities $I_s$, $M_0^2$ and $N_0^2$ of the
Josephson, interference and qua\-si\-par\-ticle currents,
respectively; (d) -- spatially averaged DOS of the bound states. }
\label{spectral} \vspace{-3mm}
\end{figure}

The behavior of the DOS calculated by numerical solution of
Eq.~(\ref{EqTheta0}) is shown in Figs.~\ref{dosex} and
\ref{dos2d}. At $\phi=0$ (we consider the phase within the period
$0\leq \phi <2\pi$), the DOS approaches the BCS energy dependence
depicted by the dashed line in Fig.~\ref{dosex}(a). The current
through the junction affects the singularity of DOS at the bulk
energy gap, which becomes a beak-shaped peak with root
singularities of the derivatives resulting from the divergence of
the decay length $k^{-1}(E)$ at $E \to \Delta$ [see
Eq.~(\ref{EqTheta})]. This divergence\cite{BV} is an analog of the
long-range proximity effect at $E \to 0$ in NS structures (see,
e.g., a review\cite{longrange}). The DOS identically turns to zero
inside an oscillating minigap, at $E <E_\ast(t) < \Delta$. Similar
conclusions can be drawn regarding behavior of the current
spectral densities shown in Fig.~\ref{spectral}(a--c).

Within the subgap region, $E_\ast(t) < E <\Delta$, the DOS
decreases at the distance $\gtrsim \xi_0$ from the barrier (more
pre\-cisely, at $\xi_0 |E/\Delta-1|^{-1/4}$, due to the long-range
proximity effect mentioned above), as shown in
Fig.~\ref{dosex}(b). This implies that the subgap states form an
oscillating cluster of bound states, ``Andreev band'', which has
tri\-angular shape in the $(E,x)$-space as shown in
Fig.~\ref{dos2d} (similar cluster appears near the lower gap edge,
$E=-\Delta$). The energy depth of the cluster is
\begin{align}
\delta E = \Delta-E_\ast \approx 6\Delta W^{4/5}
\sin^{8/5}(\phi/2),  \label{depth}
\end{align}
[see \Eq{E*IPT} below], and it spreads over the distance of
several $\xi_0$ from the tunnel barrier. We note that because of
comparatively large value of $\delta E$, we may neglect in the
spectral characteristics a smaller ($\sim \Delta W$) effect of
local suppression of the order parameter $\Delta$ in the vicinity
of the barrier.\cite{BBG} Obviously, the quantity $\delta E$ plays
the role of the smallest characteristic energy of the spectrum in
\Eq{Cond}.

The small value of the parameter $W$ enables us to apply a
perturbative approach for solving \Eq{EqTheta0}. At the energies
far enough from the gap edge, where $\theta$ is of the order of
unity, the quantity $\theta_0$ is close to $\theta_S$, and
\Eq{EqTheta0} leads to the following asymptotic relation for
$\theta_0$,
\begin{equation}
\cosh\theta_0 = \cosh\theta_S - 2\gamma(t)\sinh 2\theta_S
\bigl(i\sinh^3\theta_S \bigr)^{1/2}. \label{OPT}
\end{equation}

However, when the energy approaches $\Delta$, this expansion fails
due to divergence of $\theta_S$ at the bulk gap edge. This
requires modification of the perturbative theory within the region
$|E-\Delta| \ll \Delta$,\cite{BBG} where the quantities $\theta_0$
and $\theta_S$ are large, while their difference,
$\theta_0-\theta_S$, may have arbitrary value. Using these
arguments, we hold only large exponents, $\exp\theta_{0,S}$, in
the hyperbolic functions in the right-hand side (rhs) of
\Eq{EqTheta0} and use the asymptotic expression, $\exp\,\theta_{S}
\approx \sqrt{2\Delta/(E-\Delta)}$ at small $|E-\Delta|$. Then,
introducing a dimensionless energy variable $\epsilon$ and the
normalized spectral function $y(\epsilon)$ according to relations
\begin{subequations}\label{EqIPT}
\begin{align}
&\epsilon = p^2(t) (E - \Delta)/ {2\Delta}, \quad &p(t)=
\bigl[2/\gamma^2(t)\bigr]^{1/5}, \label{Eqy}
\\
&\exp\theta_0 = p(t)y(\epsilon), \qquad &\exp\theta_S=
p(t)\epsilon^{-1/2}, \label{expIPT}
\end{align}
\end{subequations}
we reduce \Eq{EqTheta0} to a numerical algebraic equation for
$y(\epsilon)$,
\begin{align}
\bigl(y\sqrt{\epsilon} - 1\bigr)^2 = iy^5. \label{Eqy1}
\end{align}
According to \Eqs{EqIPT}, the parameter $p(t) \sim \delta
E^{-1/2}$ determines a characteristic scale of the spectral
functions $u_0, v_0 \approx (1/2)\exp\theta_0$ in the vicinity of
the bulk gap edge.

The choice of the relevant complex root $y(\epsilon)$ of \Eq{Eqy1}
is determined by the requirement for the asymptotic behavior at
$\epsilon \gg 1$, which must coincide with the energy dependence
given by direct perturbative expansion \eqref{OPT} at $\delta E
\ll |E-\Delta| \ll \Delta$. Within the main approximation, the
function $y(\epsilon)$ turns to zero at large $\epsilon$ as
$\epsilon^{-1/2}$. At the energies $\epsilon$ smaller than
$\epsilon_\ast = -(25/6)(2/3)^{1/5} \approx -3.84$, the function
$y(\epsilon)$ becomes imaginary, and therefore the spectral
functions
\begin{align}
N_0 = M_0 = \frac{1}{2} p(t)\re y , \quad I_s = -\frac{1}{4}
p^2(t)\im y^2 \sin \phi , \label{IsIPT}
\end{align}
turn to zero at $E < E_*(t)$, where the minigap edge $E_*(t)$ is
determined by relation following from \Eq{Eqy},
\begin{equation}
E_\ast (t) = \Delta\bigl[1 - C\gamma^{4/5}(t)\bigr], \;\; C =
{25}/({3\cdot 6^{1/5}}) \approx 5.82. \label{E*IPT}
\end{equation}
We note that due to moderately small value $W=0.05$ of the
transparency parameter, the minigap values, obtained by numerical
solution of \Eq{EqTheta0} and shown in Figs.~\ref{dosex} and
\ref{spectral}, slightly differ from their approximate values
found from \Eq{E*IPT} (by some 10\%). Using the IPT approximation,
however, significantly simplifies both the analytical and
numerical calculations, allowing us to factorize the time and
energy dependencies of the spectral functions. Applying this
approximation to \Eq{SolTheta}, we obtain the spatial dependence
of the spectral angle,
\begin{align}
\exp\theta(x,E,t) &={p(t)}\epsilon^{-1/2}\tanh^2 [s(\epsilon) -
k(\epsilon,t)z/2], \label{thetaIPT}
\\
\tanh^2 s(\epsilon) &= y(\epsilon)\sqrt{\epsilon}, \quad
k(\epsilon,t) = \bigl[2\sqrt{\epsilon}/ip(t)\bigr]^{1/2}.
\label{sIPT}
\end{align}
It is interesting to note that the spectral functions are
inhomogeneous within the leads at $E=\Delta$ and finite at any
finite distance from the barrier, as it is seen in
Fig.~\ref{dos2d}, which seems to contradict the divergence of the
characteristic decay length $k^{-1}(E)$ of the spectral angle at
the gap edge. The explanation to this effect is the following: At
$E\to \Delta$ ($\epsilon \to 0$), the functions $s(\epsilon)$ and
$k(\epsilon,t)$ in \Eq{sIPT} turn to zero as $\epsilon^{1/4}$,
which cancels the divergence of the prefactor in \Eq{thetaIPT} and
results in the following expression for the spectral angle at the
gap edge,
\begin{align}\label{E=D}
\exp\theta(x,\Delta,t) = p(t)y(0) [1 - z/\sqrt{2ip(t)y(0)}]^2.
\end{align}
According to \Eq{E=D}, the spectral functions at $E = \Delta$ and
$z \to \infty$ diverge, which restores their BCS-divergence at the
gap edge within the bulk superconductor.

\section{Distribution function}

To first order in $eV$, equation \eqref{KineticG} in the Wigner
representation has the form of a diffusive kinetic
equation,\cite{LOnoneq}
\begin{equation}
N \partial_t f = {\mathcal{D}}\partial_x \left(D_+\partial_x
f\right), \label{Kin1}
\end{equation}
where $D_+ = (1/2)(1 + |u|^2 - |v|^2)$ is the dimensionless
diffusion coefficient. At the reservoir, the distribution function
approaches equilibrium value, $f(-L) = n_{\textit{eq}}(E)$. The
boundary condition at the tunnel barrier following from
\Eq{BoundaryGK},
\begin{align}
(D_+\partial_x f)_0 = \beta\partial_E f_0, \label{Bound1}
\end{align}
determines the boundary value of the flow $D_+\partial_x f$ of
nonequilibrium quasiparticles escaping from the barrier. In this
equation, the function
\begin{align}
\beta(E,t) = eV I_s(E,t) W/\xi_0
\end{align}
describes the source of the nonequilibrium - the energy exchange
between the quasiparticles and the superfluid condensate due to
the oscillations the supercurrent spectral density $I_s$. This
results in the DOS oscillations which can be expressed through
$I_s$ by integrating the identity \eqref{Equal1} over the left
lead, taking into account \Eq{Equal2} and extending the
integration over the whole negative semi-axis due to rapid decay
of the time derivatives of all quantities at a small distance $x
\sim \xi_0$,
\begin{equation}
\langle \partial_t N \rangle_x(E,t) \equiv \int^{0}_{-\infty}
\partial_t N(x,E,t)\,dx = {\mathcal{D}} \partial_E \beta(E,t).
\label{EqualN}
\end{equation}

Due to the existence of oscillating Andreev band in the vicinity
of the barrier, the behavior of the quasiparticles strongly
depends on their energy. Indeed, most of excitations with the
energy $E > \Delta$ spend a short time $\sim \Delta^{-1}$ near the
barrier, they rapidly diffuse away along the leads and escape into
the reservoirs (travelling quasiparticles). The quasiparticles
with low energy approaching the contact at the distance $\sim
\xi_0$ while the depth of the An\-dre\-ev band increases ($0 <
\phi < \pi$) are trapped by the Andreev band and spend much larger
time $\sim (eV)^{-1}$ inside the band, following the spectrum
oscillations. During the next half-period ($\pi < \phi < 2\pi$),
the depth of the Andreev band decreases, and the trapped
quasiparticles are pushed out to the extended states. Such a
physical picture is similar to the quasiparticle dynamics within
the surface skin layer of a superconductor irradiated by rf
electromagnetic field.\cite{rf} Below we will perform separate
analysis of the kinetics of the travelling and trapped
quasiparticles.

\subsection{Travelling quasiparticles}

At $E > \Delta$, the coefficients $N$ and $D_+$ in \Eq{Kin1}
rapidly vary in the near vicinity $x \sim \xi_0$ of the barrier
and then, at $x \gg \xi_0$, approach their values in a bulk
superconductor, $N \to N_S(E) = E/\sqrt{E^2 - \Delta^2}$, $D_+ \to
1$. Correspondingly, the solution of \Eq{Kin1} has the form of a
slowly varying function $f^{(0)}$ with a small but rapidly varying
addition $f^{(1)}$, which vanishes at the distances $x \gg \xi_0$
from the barrier. Thus, within the main approximation, the
population of the extended states, $f_> = f^{(0)}(x, E >
\Delta,t)$, satisfies \Eq{Kin1} with the asymptotic values of the
coefficients at $x \gg \xi_0$,
\begin{equation}
N_S \partial_t f_> = {\mathcal{D}}\partial^2_x f_>. \label{Eqf>}
\end{equation}

To derive effective boundary condition to this equation, we
subtract \Eq{Eqf>} from the initial kinetic equation \eqref{Kin1},
keeping large spatial derivative of the function $f^{(1)}$,
\begin{equation}
(N - N_S)\,\partial_t f_> ={\mathcal{D}}\partial_x \bigl[(D_+ -1)
\partial_x f_> +  D_+\partial_x f^{(1)}\bigr], \label{Bound3}
\end{equation}
and then integrate \Eq{Bound3} over $x$ along the left lead. Since
all terms in this equation vanish at $x \gg \xi_0$, we extend the
in\-te\-gration over the whole semi-axis, similar to \Eq{EqualN},
and take the smooth function $f_>$ in its left-hand side (lhs) at
$x=0$. Then, using \Eq{Bound1}, we obtain the boundary condition,
\begin{equation}
\langle N - N_S\rangle_x (\partial_t f_>)_0  =
{\mathcal{D}}\left(\beta
\partial_E f_> - \partial_x f_> \right)_0. \label{Bound>}
\end{equation}

The distribution function $f_>$ consists of a static part $g(x,E)$
and a dynamic (time-dependent) part, $f_>(x,E,t) = g(x,E)+
h(x,E,t)$. The static part linearly varies along the lead and
approaches equilibrium distribution function $n_{\textit{eq}}(E)$
at $x=-L$,
\begin{equation}
g(x, E) = g_0(E) + (x/L)[g_0(E) - n_{\textit{eq}}(E)].
\label{Solg}
\end{equation}
The dynamic component $h(x,E,t)$ has the form,
\begin{align}
&h(x,E,t) = \sum\nolimits_{n} h_n(E)e^{-i neV t + K_n x}, \;\;
\langle h \rangle_t=0, \label{Solh}
\\
&K_n(E) = \sqrt{|n|eV N_S(E){\mathcal{D}}^{-1}}
\exp\left[-(i\pi/4) \sgn n\right].\nonumber
\end{align}
Here $\langle \ldots \rangle_t$ denotes time averaging over the
period $\tau = 2\pi/eV$, and $h_n(E)$ are the Fourier harmonics of
the function $h(0,E,t)$,
\begin{equation} \nonumber
h_n(E) = \int^{\tau}_{0} \frac{dt}{\tau} h(0,E,t) e^{ineVt} .
\nonumber
\end{equation}
Equation \eqref{Solh} was obtained assuming the decay length $L_V
= K_n^{-1} \sim \sqrt{\mathcal{D}/eV}$ of the oscillations of the
distribution function to be much smaller than the junction length
$L$; this enables us to apply the solution for a semi-infinite
lead.  In the opposite case, $L_V \gg L$, the reservoirs
effectively suppress these oscillations, and the function $h$
rapidly decays while the voltage decreases. The equation for the
harmonics $h_n(E)$ follows from \Eq{Bound>},
\begin{align} \label{Eqhn1}
&{\mathcal{D}}^{-1}\langle N - N_S \rangle_x \partial_t h(0,E,t) -
\beta
\partial_E h(0,E,t) + \left(\partial_x g\right)_0
\nonumber\\
&+\sum\nolimits_{n} h_n(E) K_n e^{-ineVt} = \beta\,
\partial_E g_0(E).
\end{align}
The first two terms in the lhs of \Eq{Eqhn1}, proportional to
small applied voltage $eV$, can be neglected as compared to the
fourth term. Then, averaging \Eq{Eqhn1} over $t$ and using the
identity \eqref{EqualN}, we obtain the expression for the third
term,
\begin{equation}
\left(\partial_x g\right)_0  = \partial_E \langle \beta h
\rangle_t, \label{third}
\end{equation}
which therefore is also proportional to small $eV$ and can be
neglected as well. As the result, the approximate solution of
\Eq{Eqhn1} reads $h_n = \beta_n K_n^{-1} \partial_E g_0$, i.e.,
the dynamic part $h(x,E,t)$ of the distribution function is
expressed through the boundary value of the static part, $g_0(E)$.

It follows from this consideration that the dynamic part of the
distribution is generated by the oscillations of the quasiparticle
spectrum at $E > \Delta$. These oscillations provide the energy
transfer from the electric field to quasiparticles, which
generally results in the `heating' of quasiparticles, i.e., their
redistribution towards higher energies (pumping effect). The
heating is described by a diffusion equation for $g_0$ in the
energy space following from the \Eqs{third} and \eqref{Solg},
\begin{align}
\partial_E &\left(D_E\,
\partial_E g_0\right) = L(\partial_x g)_0 = g_0 - n_{\textit{eq}},
\label{Solg0}
\\ \nonumber
D_E &= L\sum\nolimits_{n} |\beta_n|^2 K_n^{-1}.
\end{align}
Similar equation has been obtained in Refs.~\onlinecite{NoiseSNS}
for MAR-in\-duced heating in long SNS junctions, where the
nonequilibrium is constrained by inelastic collisions. In our
strongly inhomogeneous case, the role of relaxation factor in
\Eq{Solg0} is played by the diffusive flow $(\partial_x g)_0$ of
nonequilibrium quasiparticles from the junction. The intensity of
the heating effect is determined by the estimate of the diffusion
term in \Eq{Solg0} with the equilibrium function
$n_{\textit{eq}}$, $\partial_E\left(D_E
\partial_E n_{\textit{eq}}\right) \sim (L/\xi_0)
W(eV/\delta E)^{3/2}$. Due to the presence of the two small
factors, this term is small for reasonable lengths of the leads,
which enables us to approximate the function $g_0$ with the
equilibrium distribution function, $g_0\approx n_{\textit{eq}}$.
Then the boundary value of the distribution function of travelling
quasiparticles reads,
\begin{align}
f_>(0, E, t) &= n_{\textit{eq}}(E) + h(0, E, t),
\\
h(0, E, t) &=  n_{\textit{eq}}' \sum\nolimits_{n}
\beta_{n}K_n^{-1} e^{-ineV t},  \;\; n_{\textit{eq}}' \equiv
\partial_E n_{\textit{eq}}(E). \label{f>}
\end{align}

\subsection{Trapped quasiparticles}

At small voltages, $eV \ll \Delta$, the spatial size $\xi_0$ of
the Andreev band is much smaller than the smallest kinetic length
$L_V$, therefore the main part $f^{(0)}$ of the distribution
function of the trapped quasiparticles is spatially uniform, $f_<
\approx f^{(0)}(E<\Delta,t)$. Then the kinetic equation
\eqref{Kin1} takes the form,
\begin{equation}
 N\,\partial_t f_< = {\mathcal{D}}
\partial_x\bigl(D_+ \partial_x f^{(1)} \bigr).\label{Eqf<1}
\end{equation}
By averaging \Eq{Eqf<1} over $x$ and using the boundary condition
\eqref{Bound1}, we obtain partial differential equation (PDE) for
the function $f_<$,
\begin{equation}
\langle N \rangle_x\partial_t f_< - {\mathcal{D}}\beta\partial_E
f_< =0. \label{PDE}
\end{equation}
As long as the coefficients of this PDE satisfy the identity
\eqref{EqualN}, the equation $d\xi=\langle N \rangle_x \, dE +
{\mathcal{D}}\beta\, dt =0$ for its characteristic curves
$\xi(E,t) = \const$ in the $(E,t)$ space is easily integrated,
\begin{equation} \label{xi}
\xi(E, t) = \xi_0^{-1} \int^{E}_{E_*(t)}dE' \bigl\langle N(x,E',
t) \bigr\rangle_x.
\end{equation}
This allows us to reduce the PDE \eqref{PDE} to the ordinary
differential equation for the function $f_<$ along the
characteristics,
\begin{equation}
\partial_t f_<(\xi, t)|_{\xi=\const} = 0. \label{Eqfxi}
\end{equation}
The quantity $\xi(E,t)$ has the meaning of averaged number of
states with the energy smaller than given energy $E$, normalized
over $N_F$; in the bulk superconductor, it approaches
$\Theta(E-\Delta) \sqrt{E^2-\Delta^2}$. The bottom $E_\ast(t)$ of
the Andreev band represents the reference point,
$\xi[E_\ast(t),t]=0$. Within the IPT approximation, \Eq{xi} reads
\begin{subequations} \label{Jn}
\begin{align}
\xi(\epsilon,t) &=  \Delta [4\gamma(t)]^{1/5}J(\epsilon), \quad
J(\epsilon) = \int_{\epsilon_\ast}^\epsilon d\epsilon'
n(\epsilon'), \label{J}
\\
n(\epsilon) &= |\epsilon|^{-1/2}\im \sqrt{ iy(\epsilon)},
\label{n}
\end{align}
\end{subequations}
where the function $n(\epsilon)$ is related to the averaged DOS as
$\langle N \rangle_z = [2\gamma^3(t)]^{-1/5}n(\epsilon)$. As
follows from \Eq{n}, the function $\langle N \rangle_z(E)$ plotted
in Fig.~\ref{spectral}(d) exhibits a singularity at the bulk gap
edge, due to the long-range proximity effect in the superconductor
[see comments to \Eq{EqTheta0}].

\begin{figure}[tb]
\epsfxsize=8.5cm\epsffile{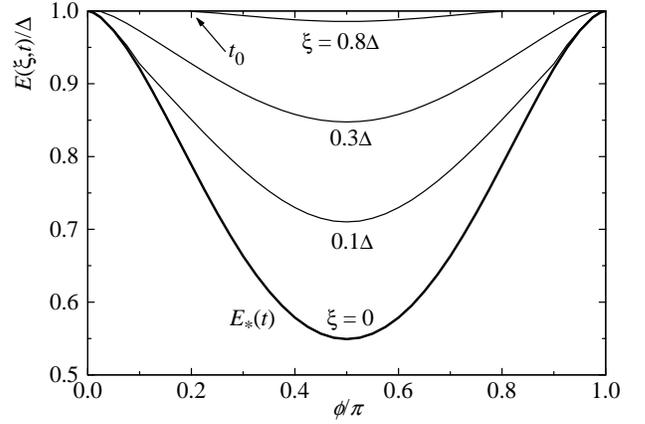} \vspace{-2mm}
\caption{Levels of equal number of states $\xi(E,t)=\const$ vs
phase difference at $W=0.05$. Bold line depicts the edge
$E_\ast(t)= E(\xi=0,t)$ of the minigap, and $t_0$ indicates a
particular moment of trapping to the trajectory $\xi=0.8\Delta$.}
\label{levels} \vspace{-3mm}
\end{figure}

According to \Eq{Eqfxi}, the distribution of trapped excitations is a
function of the ``integral of motion'' $\xi(E, t)$,
\begin{equation}
f_<(E,t) = f[\xi(E, t)], \label{fxi}
\end{equation}
and therefore holds a constant value along the trajectories
$\xi(E,t) = \const$ in the $(E,t)$ space shown in
Fig.~\ref{levels}. These trajectories can be interpreted as
diffusive analogue of Andreev levels adiabatically moving in a
slowly varying external field (cf. Ref.~\onlinecite{Arg1}). From
this we conclude that the trapped quasiparticles are completely
dragged by the spectrum oscillations, and their energy
periodically changes, in accordance with the relation following
from \Eqs{xi} and \eqref{EqualN},
\begin{equation}
\partial_t E |_{\xi = \const} = -{\mathcal{D}} \beta \langle N\rangle_x^{-1}
. \label{dEdt}
\end{equation}

Since the bound states at the tunnel barrier appear and vanish
periodically, their population is imposed by the distribution of
travelling quasiparticles at the bulk gap edge. This is expressed
by the matching condition
\begin{equation}
f_<(\xi)= f_>[0, \Delta, t_0(\xi)], \;\; 0<\phi
 < \pi. \label{Bound<>}
\end{equation}
Here $t_0(\xi)$ is the moment of the crossing the gap edge by
quasiparticle with the given value of $\xi$ (see Fig.~\ref{levels}),
i.e., it is the smallest solution of the equation $E(\xi,t_0)=\Delta$
within the period of the DOS oscillations.

Since the trapped quasiparticles spend long time at the barrier,
the heating effect for them becomes well pronounced, in contrast
to the travelling quasiparticles. Indeed, as follows from \Eq{f>},
the time-dependent part $h$ of the distribution function turns to
zero at $E =\Delta$ due to singularity of $N_S$, and therefore the
distribution function of trapped quasiparticles turns into a
plateau (i.e., does not depend on energy),
\begin{equation}
f_<(\xi) = g_0(\Delta) = n_{\textit{eq}}(\Delta). \label{plateau}
\end{equation}
%

Non-adiabatic correction to the distribution function
\eqref{plateau} is produced by inelastic relaxation. To include
the effect of inelastic collisions on the quasiparticle kinetics,
we will add a collision term in the $\tau$-appro\-xi\-mation,
$N\tau_\epsilon^{-1}[n_{\textit{eq}}-f]$, to the rhs of the
kinetic equation \eqref{Eqf<1}. Within such a model, the kinetic
equation for the trapped quasiparticles reads
\begin{equation} \label{Eqtau1}
\partial_t f_<(\xi,t) = \tau_\epsilon^{-1}\bigl\{n_{\textit{eq}}[E(\xi,t)]-
f_<(\xi,t)\bigr\}.
\end{equation}
The initial condition to \Eq{Eqtau1} follows from \Eq{Bound<>},
\begin{equation} \label{Btau1}
f_<(\xi,t_0) = n_{\textit{eq}}(\Delta).
\end{equation}

\section{Electric Current}

Proceeding to the calculation of the electric current \eqref{I},
we start from the analysis of the ac Josephson current
$I_1\sin\phi$. First, we note that the effect of the DOS
oscillations on the amplitude $I_1$, calculated with the
equilibrium distribution function $n_{\textit{eq}}$, gives rise
only to small corrections to the tunnel model result, $I_1^{\it
eq} = (\pi\Delta/{2eR}) \tanh ({\Delta}/{2T})$.\cite{Wert,LO}
Indeed, in this case, the integral in \Eq{I1} is determined by the
imaginary Matsubara energies $i\omega_n = \pi i T(2n+1)$ far from
the gap edges, which allows us to apply ordinary perturbation
approach [see \Eq{OPT}] to the calculation of the spectral density
$I_s$. This results in the correction of the order of $W$ to
$I_1^{\it eq}$,\cite{BBG} which will be neglected below.

In what follows, we will focus on the nontrivial contributions of
nonequilibrium quasiparticles to the Jo\-seph\-son current $I_1$.
Such contributions, $I_h(t)$ and $I_<(t)$, come from the dynamic
part $h$ of the distribution of travelling quasiparticles and the
distribution $f_<$ of trapped quasiparticles,
\begin{subequations} \label{dI1}
\begin{align}
I_h(t) &= \frac{1}{eR}\sum_{n\neq 0} {e^{-ineV t}}
\int_\Delta^\infty   dE \, I_s K_n^{-1}\beta_n n_{\textit{eq}}'(E)
, \label{I1h}
\\
I_<(t) &= \frac{1}{eR} \int_{E_\ast}^\Delta dE\, I_s
\{f_<[\xi(E,t),t]\, -\, n_{\textit{eq}}(E)\}. \label{I1<}
\end{align}
\end{subequations}

Due to rapid convergence of the integral in \Eq{I1h} at $E -\Delta
\lesssim \delta E \ll (\Delta, T)$, the function
$n_{\textit{eq}}'(E)$ can be taken at $E=\Delta$.  For similar
reason, we apply the IPT expression \eqref{IsIPT} for the spectral
density $I_s$, which leads to the following result,
\begin{equation}\label{It}
I_h(t) = \frac{\Delta F_1 }{eR}  \sqrt{\frac{eV}{2\Delta}}\,
W^{2/5} P(t)\sin\phi,
\end{equation}
where
\begin{align}
\nonumber F_1(T) &= \Delta n_{\textit{eq}}'(\Delta) =
\frac{\Delta}{2T} \cosh^{-2} \frac{\Delta}{2T},
\\
\nonumber P(t) &= C_1 \sin^{2\nu} \frac{\phi}{2} \sum_{m=1}^\infty
\frac{(-1)^m \sqrt{m} \sin (m\phi -\pi/4)}{(m^2 -\nu^2)
\mathop{\mathrm{B}}(\nu+m, \nu-m)},
\\ \nonumber
C_1 &= 2^{1-\nu/2} \int_0^\infty d\epsilon\, \epsilon^{1/4} [\im
y^2(\epsilon)]^2 \approx 0.15,\;\;\; \nu=0.2,
\end{align}
and $\mathop{\mathrm{B}}(x,y)$ is the Euler's beta function. The time average
of the current (\ref{It}) does not vanish,
\begin{align}
I_h^{\it dc} &\equiv \langle I_h  \rangle_t = \frac{\Delta
F_1}{eR} C_3 W^{2/5} \sqrt{\frac{eV}{2\Delta}}, \label{<IJ>}
\\
C_3 &= \langle P(t)\sin\phi \rangle_t = 0.018, \nonumber
\end{align}
and, moreover, it exhibits a strongly nonlinear ($\sim \sqrt V$)
voltage dependence. We recall that the validity of these results
is restricted by the condition of small diffusion length as
compared to the junction length, $L_V = \sqrt{\mathcal{D}/eV} \ll
L$, or, equivalently, $eV \gg E_{\textit{Th}}$, where
$E_{\textit{Th}} = \mathcal{D}/L^2$ is the Thouless energy [see
comments to \Eq{Solh}]. In the opposite case, $eV \ll
E_{\textit{Th}}$, the oscillations of the distribution function
are damped by equilibrium reservoirs, and the nonequilibrium dc
current, produced by travelling quasiparticles, rapidly vanishes.
Similar damping effect is caused by inelastic collisions, when the
inelastic scattering length becomes smaller than the diffusion
length $L_V$.

The dissipative dc current of travelling quasiparticles results
from non-adiabatic (diffusive) evolution of the distribution
function $h(t)$. Similar conclusion is also true for the trapped
quasiparticles: only the non-adiabatic part of the distribution
function contributes to the dissipative dc current. Indeed,
inserting the adiabatic distribution function
$n_{\textit{eq}}(\Delta)$, \Eq{plateau}, into \Eq{I1<}, and taking
advantage of a small integration interval, which allows us to
expand the difference of the equilibrium functions over
$E-\Delta$, we get,
\begin{align}\label{Iad<}
I_<^{\textit ad}(t) &= \frac{\Delta F_1}{eR} C_2  W^{4/5}
\sin^{8/5} \frac{\phi}{2} \sin \phi,
\\
C_2 &= 2^{-2\nu} \int_{\epsilon_\ast}^0 d\epsilon \, \epsilon \im
y^2(\epsilon) \approx 3.11. \nonumber
\end{align}
This equation contains only odd harmonics of the ac Josephson
current. The dc current results from the non-adiabatic correction
$\delta f$ to the distribution function, found from the solution
of the kinetic equation \eqref{Eqtau1},
\begin{align} \label{Ieps}
I_<^{\textit dc} = \frac{1}{eR} \Bigl\langle \int_{E_\ast}^\Delta
dE\, I_s\, \delta f[\xi(E,t),t] \Bigr\rangle_t \!\!= \frac{\Delta
F_1}{eR} W^{4/5} K(eV).
\end{align}
The factor $K(eV)$ in \Eq{Ieps} is a complicated numerical
function of the relaxation parameter $eV \tau_\epsilon$. In the
limits of weak ($eV \tau_\epsilon \gg 1$) and strong ($eV
\tau_\epsilon \ll 1$) relaxation, this function decreases as $(eV
\tau_\epsilon)^{-1}$ and $eV \tau_\epsilon$, respectively. In the
spirit of our modelling approach to the problem of inelastic
scattering, it is reasonable to approximate $K(eV)$ by usual
relaxation factor
\begin{equation} \label{Kapp}
K(eV) \approx \frac{\alpha_1 eV \tau_\epsilon}{\alpha_2+ (eV
\tau_\epsilon)^2},
\end{equation}
where the parameters $\alpha_{1,2}$ are to be evaluated from the
asymptotics of $K(eV)$ in both limiting cases.

In the weak relaxation limit, $eV \tau_\epsilon \gg 1$, the
adiabatic distribution function is given by \Eq{plateau},
$f_<^{\it ad} = n_{\textit{eq}}(\Delta)$; then the non-adiabatic
correction $\delta f$ is determined by the local energy $E(\xi,t)$
averaged along the trajectories $\xi = \const $,
\begin{equation}\nonumber 
\delta f(\xi,t) =  \tau_\epsilon^{-1} n_{\textit{eq}}'(\Delta)
\int_{t_0(\xi)}^t dt' \, [E(\xi,t')-\Delta].
\end{equation}
Substituting $\delta f$ into the expression \eqref{Ieps} and using
the IPT relations \eqref{Jn} for the function $\xi(E,t)$, we
obtain the coefficient $\alpha_1$ in the relaxation factor
$K(eV)$,
\begin{align} \nonumber
\alpha_1 = \frac{ \mathrm{B} (0.5,2.3)} {2^{2/5} \pi}
\int_{\epsilon_\ast}^0 d\epsilon \im y^2(\epsilon) \int_0^1 dx\,
\epsilon_0 \bigl[x^{1/5} J(\epsilon) \bigr] \approx 1.38,
\end{align}
where $\epsilon_0(\zeta)$ is the solution of equation $\zeta =
J(\epsilon_0)$, and the function $J(\epsilon)$ is defined in
\Eq{J}.

In the opposite limit of strong relaxation, $eV\tau_\epsilon \ll
1$, the initial condition \eqref{Btau1} is quickly `forgotten',
and the adiabatic part of the distribution is the equilibrium
function of the local energy, $f^{\it ad}_< =
n_{\textit{eq}}[E(\xi,t)]$. In this case, the non-adiabatic
addition $\delta f$ is proportional to the time derivative of
$E(\xi,t)$ along the trajectory,
\begin{equation} \\ \nonumber 
\delta f(\xi,t) \approx -\tau_\epsilon n_{\textit{eq}}'(\Delta)\,
\partial_t E(\xi,t).
\end{equation}
Using the expression \eqref{dEdt} for $\partial_t E$ and
calculating $I_<^{\textit dc}$ in \Eq{Ieps} in the IPT
approximation, we arrive at the following relation between the
coefficients $\alpha_1$ and $\alpha_2$,
\begin{equation} \label{a1a2}
\frac{\alpha_1}{\alpha_2} = \int_0^\pi \frac{d\tau\, \sin^2
2\tau}{2^{7/5}\pi\sin^{2/5}\tau} \int_{\epsilon_\ast}^0  d\epsilon
\, \sqrt{|\epsilon|} \frac{ [\im y^2(\epsilon)]^2}{\im
\sqrt{iy(\epsilon)}}\approx 1.74. \nonumber
\end{equation}

Now we turn to evaluation of the quasiparticle current $I_2$ and
the interference current $I_3\cos\phi$. These currents are
proportional to small applied voltage, and therefore they can be
calculated within the equilibrium approximation for the
quasiparticle distribution neglecting small nonequilibrium
correction. The contribution of the trapped quasiparticles is
small, because of the small occupied phase volume, and can be
neglected. Moreover, in the limit of weak relaxation, this
contribution identically turns to zero because of
energy-in\-de\-pen\-dent distribution $f_<$. Within the BCS
approximation for the spectral functions $N_0$ and $M_0$, the
integrals in \Eqs{I2} and \eqref{I3} logarithmically diverge at
the gap edge. The effect of the phase difference on the
quasiparticle spectrum eliminates this divergence and effectively
cuts the spectral functions at the value $(\delta E)^{-1/2}$ [see
comments to \Eq{EqIPT}], which is equivalent to the effective
cut-off in energy, $E -\Delta \gtrsim \delta E$. Using the IPT
expressions \eqref{IsIPT} for the spectral functions at $E-\Delta
\ll \Delta$, we obtain to the main order in $W$,
\begin{align}
I_2(t) &= I_3(t) + \frac{V}{R}[1- n_{\textit{eq}}(\Delta)],
\label{ResI2}
\\
I_3(t) &= \frac{V}{R} \Bigl[ F_1 \ln
\sqrt{a\bigl/{\gamma^{4/5}(t)}} -F_2 \Bigr], \label{ResI3}
\end{align}
where
\begin{align}
F_2(T) &=\int_\Delta^\infty \frac{\Delta^2 dE}{E^2-\Delta^2}
\bigl[n_{\textit{eq}}'(\Delta)-n_{\textit{eq}}'(E)\bigr],
\nonumber
\\ \nonumber
\ln a &= \int_0^\infty d\epsilon \Bigl[\re^2 y(\epsilon) -
\frac{\Theta(\epsilon-1)}{\epsilon}\Bigr]+ \frac{2}{5} \ln 2 \approx
0.13.
\end{align}

We note that within the tunnel model,\cite{Wert,LO} the role of
the cut-off factor is played by the quantity $eV$, which enters
large logarithmic factor in the expression for the current. In our
case, according to the adiabaticity criterion \eqref{Cond}, the
effective cut-off factor, $\gamma^{4/5} \sim \delta E$, is much
larger. From this we conclude that within the region of
applicability of the adiabatic approach, $eV \ll \delta E$, the
quasiparticle and the interference currents are logarithmically
smaller as compared to the results of the tunnel model. While the
voltage increases and exceeds the depth of the Andreev band, $eV >
\delta E$, the tunnel model approximation for the currents $I_2$
and $I_3$ becomes valid.

Due to the presence of a time-de\-pen\-dent factor $\gamma(t)$ in
the logarithmic terms, the currents $I_2$ and $I_3$ oscillate in
time and exhibit logarithmic singularities when $\gamma(t)$ turns
to zero, i.e., at $\phi = 0$. In the vicinity of these points, the
Andreev band vanishes, which violates the condition of
adi\-aba\-ti\-ci\-ty \eqref{Cond}. Following remarks to the
equations \eqref{ResI2} and \eqref{ResI3}, we can
semi-quantitatively describe the whole shape of the oscillations
of the quasiparticle and interference currents by adding a small
regularization term $eV/\Delta$ to the denominator of the
logarithmic argument in \Eq{ResI3}.

\begin{figure}[tb]
\epsfxsize=8.3cm\epsffile{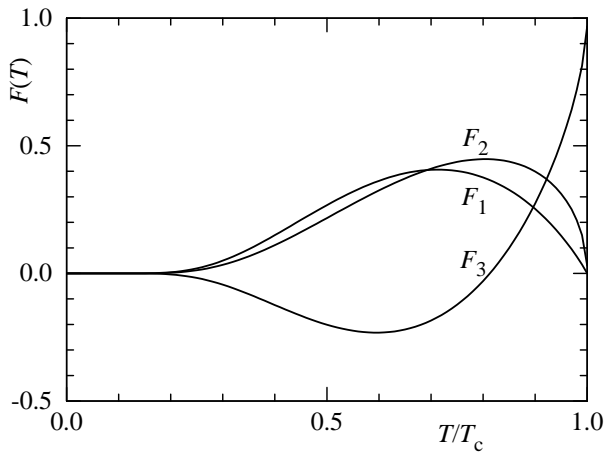} \vspace{-2mm}
\caption{Temperature dependencies of the functions $F_1$, $F_2$,
and $F_3$. }
\label{f} \vspace{-3mm}
\end{figure}

By averaging \Eq{ResI2} over time, we obtain the dissipative part
of the current $I_2(t)$,
\begin{align}  \nonumber
\langle I_2 \rangle_t &= \frac{V}{R} \Bigl[ F_1 \ln \sqrt{
b/W^{4/5}} + F_3\Bigr] , 
\\ \nonumber
\ln b &= \ln a + \int_0^\pi \frac{d\tau}{\pi} \ln \sin^{-8/5}\tau
= 1.24,
\\ \nonumber
F_3(T) &=1- n_{\textit{eq}}(\Delta)-F_2(T).
\end{align}
The interference current also has a dc component due to the
oscillations in $I_3$, though its magnitude,
\begin{align} \nonumber
&\langle I_3 \cos\phi \rangle_t = \frac{V}{R} F_1 \ln \sqrt{c},
\\
\nonumber &\ln c = \int_0^\pi \frac{d\tau}{\pi} \cos 2\tau\, \ln
\sin^{-8/5}\tau = 0.8,
\end{align}
is small with respect to the logarithmic term in $\langle I_2
\rangle$. The sum of the quasiparticle and interference dc
currents is given by expression,
\begin{align} \label{I23dc}
I_{23}^{\it dc} \equiv \langle I_2  +I_3 \cos\phi \rangle_t =
\frac{V}{R} \bigl[ F_1 \ln  ({2.77}/W^{2/5}) +F_3\bigr].
\end{align}
The temperature dependencies of the functions $F_{1}(T)$,
$F_{2}(T)$ and $F_3(T)$ are shown in Fig.~\ref{f}.

\section{Discussion}

\begin{figure}[tb]
\epsfxsize=8.5cm\epsffile{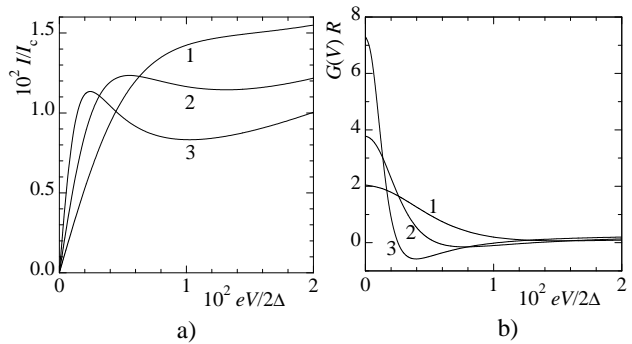} \vspace{-2mm}
\caption{dc current (a) and differential conductance (b) vs
voltage for different values of $\tau_\epsilon \Delta =50$ (1),
$100$ (2), and $200$ (3). All curves are plotted at $W=0.05$,
$T=0.5T_c$. }
\label{discurr} \vspace{-5mm}
\end{figure}

Analyzing different calculated current components, we conclude
that the most significant deviation from the results of the tunnel
model comes from the nonequilibrium Josephson current, while the
quasiparticle and interference currents undergo minor
(logarithmic) changes. There are the two features to be mentioned.
The first is a non-monotonous voltage dependence of the current of
trapped quasiparticles described by the relaxation factor $K(eV)$
[see \Eqs{Ieps} and \eqref{Kapp}], which has a maximum at $eV \sim
\tau_\epsilon^{-1}$. Due to rather large value of the parameter
$\tau_\epsilon \Delta \gtrsim 10^2$ in most superconductors, this
current may exceed linear contributions of the quasiparticle and
interference currents at moderately small values of the
transparency parameter $W$. In this case, the $I$-$V$
characteristic exhibits an ${\cal N}$-like feature, as shown in
Fig.~\ref{discurr}(a), and the linear conductance at small bias
considerably (by the factor $W^{4/5}\tau_\epsilon \Delta$) exceeds
the tunnel model conductance, which is relevant for larger bias
[zero bias conductance peak, Fig.~\ref{discurr}(b)].

The second feature is a nonlinear, proportional to $\sqrt{V}$,
contribution to the current-voltage characteristics produced by
the travelling quasiparticles. We note, however, that because of
small numerical value of the constant $C_3$ in \Eq{<IJ>}, the
crossover of $I(V)$ to nonlinear behavior for reasonable values of
$W$ actually occurs at very small voltage. The reason for this is
a numerically small magnitude of the supercurrent spectral density
$I_s(E)$ above the bulk gap edge, as is obvious from
Fig.~\ref{spectral}(a). Physically, this is due to a rapid
decrease of the probability of the ``over-the-barrier''
An\-d\-reev reflection (reflection at the energy outside the
energy gap). As the result, the density of nonequilibrium
quasiparticles produced by the oscillations of $I_s$ at $E>\Delta$
appears to be small as compared with the ``natural'' width $\delta
E$ of the perturbed spectral region, which results in smaller
contribution of the travelling quasiparticles compared to the
trapped ones.

In order to estimate the characteristic parameters of the junction
for the transparency parameter $W=0.05$ accepted in our numerical
calculations, we will assume the junction area to be $200 \times
200$ nm and the thickness of the leads as well as the elastic
scattering length to be $50$ nm. For Al leads, this results in the
sheet resistance $R_{\Box} \approx 0.3$ $\Omega$ and
$R(\xi_0)\approx 0.45$ $\Omega$ at $\xi_0 \approx 300$ nm. Then,
according to \Eq{W}, the tunneling probability, the junction
resistance and the critical current approach the values $\Gamma
\approx 0.01$, $R \approx 4.5$ $\Omega$, and $I_c \approx 70$
$\mu$A, respectively. Thus, the characteristic voltage region, at
which the nonequilibrium dc current dominates, is of the order of
several microvolts.

It follows from the presented estimates and also from
Fig.~\ref{discurr} that the nonequilibrium effect appears at
rather small applied voltage, which makes it difficult to observe
in practice because of the jumps to the Josephson branch. To
facilitate the observation, the net Josephson current must be
suppressed by applying magnetic field or using the dc SQUID setup.
Remarkably, the effect survives even in the absence of the net
Josephson current: Local Josephson currents flowing through
different junctions (or parts of the junction) generate
dissipative dc current flowing in the same direction, which is
determined by the applied voltage (i.e., time derivative of the
phase) rather than the local values of the phase difference.
Indeed, let us consider two junctions connected in parallel (dc
SQUID) and apply half-integer magnetic flux. Then the phase
differences at the junctions will be shifted by $\pi$. This will
lead to different signs of the current spectral density in
\Eq{Ieps} for different junctions, however, the nonequilibrium
distribution function will also change the sign, which can be
proven by direct calculation. Actually, the equality of dc
currents in both junctions follows from a simple fact that the
constant phase shift is equivalent to the change of the time
reference point, which obviously cannot affect the value of the
time-averaged (dc) current. Thus, the net dc current will double
while the ac current disappears. Such situation resembles the case
of a long SNS junction,\cite{longSNS,Arg1} where the Josephson
effect is suppressed due to decay of the superconducting
correlations inside the normal part of the junction, and the
nonequilibrium effect can be observed at very small applied
voltage.

It is instructive to compare the mechanism of non\-equi\-lib\-rium
tunnel current in diffusive junctions with mesoscopic picture of
superconductive tunneling in point contacts given by the MAR
theory.\cite{Bratus95,Averin95,Bratus97} It has been already
mentioned in the Introduction, that the Andreev band is the
qualitative analog of the Andreev bound levels in point contacts,
the both are formed by the same physical mechanism -- Andreev
reflection by the jump of the superconducting phase at the
junction. The time oscillations of the Andreev levels in voltage
biased point contacts, and the quasiparticle exchange between the
levels and the continuum is the adiabatic description in the time
domain of the coherent MAR.\cite{Bratus97} The latter also include
the quasiparticle transitions across the energy gap, which in the
real-time picture correspond to the Landau-Zener tunneling between
the Andreev levels (bands).\cite{Averin95} Such tunneling provides
the flow of quasiparticles along the energy axis (spectral flow),
proportional to the gradient of the distribution function in the
energy space, $\partial_E f$, within the subgap region,
$|E|<\Delta$. These processes are weak at small voltage compared
to the continuum-bound level exchange, and they are neglected
within the adiabatic approximation adopted in the present paper.
For this reason, the spectral flow of the trapped quasiparticles
is blocked, which results in a flat (energy-independent)
distribution \eqref{plateau} of these quasiparticles in the
absence of inelastic relaxation.

However, there is considerable quantitative difference between the
point contacts and diffusive junctions: In the point contacts,
most of the quasiparticles reflected by the tunnel barrier escape
to the reservoir, while in the long diffusive lead the
quasiparticles multiply collide with the barrier due to the
impurity backscattering. This essentially increases the
probability of the coherent tunneling and Andreev
reflection\cite{multiple} and, correspondingly, enhances the
effect of the phase difference on the junction spectrum. As the
result, the depth of the Andreev band in our long-arm geometry,
$\delta E\sim \Delta W^{4/5}$ in \Eq{depth}, considerably exceeds
the depth $\sim \Delta\Gamma$ of the Andreev level in a point
contact with comparable transparency.

It is interesting to mention that generally the role of the
junction transparency in tunnel junctions with diffusive
electrodes is played by the parameter $W$ rather than the barrier
trans\-parency $\Gamma$.\cite{Kupr,BBG} Indeed, comparison of the
right- and left-hand sides of \Eqs{Boundary2} shows that as long
as the `natural' scale of the currents $\hat{J}$ and $\hat{J}^K$
is proportional to the gradients of the Green's and distribution
functions in the vicinity of the barrier, $\sim \xi_0^{-1}$,  it
is the magnitude of $W$ (not $\Gamma$ itself) which determines the
effective barrier strength. For example, at large $W \gg 1$, the
commutators in \Eqs{Boundary2} are to be small, which implies
continuity of all functions at the barrier. From this we conclude
that at $W \gg 1$, the phase and voltage are continuously
distributed along the whole structure, and the barrier does not
affect the current transport even if its transparency is small,
$\Gamma \ll 1$, provided $\Gamma \gg \ell/\xi_0$. In other words,
at large $W$, the critical current of the tunnel junction,
formally estimated by the Ambegaokar-Baratoff formula for a
low-transparent junction,\cite{AB} exceeds the critical current of
diffusive superconducting leads. Physically, this effective
`blooming' of an opaque barrier results from the multiple coherent
backscattering of quasiparticles by impurities mentioned
above.\cite{multiple}

\section{Conclusions}

We have studied the current transport through the voltage biased
tunnel junction with long diffusive superconducting leads at low
applied voltage, $eV \ll 2\Delta$, and finite temperatures. In
contrast to the tunnel model,\cite{Wert,LO} we consider the
non\-equilibrium effects in the spectrum and distribution of
quasiparticles in the vicinity of the barrier. Using the small
value of the Josephson frequency with respect to other
characteristic energies of the problem, we apply a quasi-adiabatic
approach to the analysis of the time evolution of quasiparticles.
Within such approach, we obtain a physically transparent picture
of the quasiparticle spectrum adiabatically following the
time-dependent difference of the superconducting phases. This
results in local oscillations of the spectrum of travelling
quasiparticles ($E > \Delta$) and formation of oscillating Andreev
bound states (Andreev band) within the subgap region, $E_\ast(t) <
E < \Delta$, at the distance of the order of the coherence length
$\xi_0$ near the barrier. The quasiparticles trapped by the
Andreev band are completely dragged by oscillations of the
junction spectrum which reflects complete multiple Andreev
reflection (MAR) of the subgap quasiparticles and results in
generation of higher odd harmonics of the ac Josephson current.
The inelastic relaxation of the trapped quasiparticles produces a
non-adiabatic component of their distribution which manifests
itself in the nonequilibrium contribution to the dc current. At
low enough voltages, this contribution may considerably exceed the
quasiparticle dc current given by the tunnel model; by this
reason, the resulting $I$-$V$ characteristic shows ${\cal N}$-like
feature, with the maximum at $eV \sim \tau_\epsilon^{-1}$. The
travelling quasiparticles also deviate from equilibrium due to
partial drag (which corresponds to the over-the-barrier MAR)
confined by their fast diffusion from the barrier. This results in
the additional contribution to the dc current, proportional to
$\sqrt{V}$. We note that our approach can be easily extended to
the current bias regime; in this case, arbitrary time-dependent
phase must be assumed in the calculation.

Effect of travelling quasiparticles is interesting, in principle
it dominates at small voltage because of the non-analytical
voltage dependence. Unfortunately, this contribution is
numerically small in the case of the tunnel barrier considered
here, because of the small spectral density of the Josephson
current above the gap. However, there are no fundamental reasons
to expect this contribution to be small in other junctions. It
would be interesting to find favorable junction geometry.

We acknowledge useful discussion with L.~Kuzmin and D.~Winkler.
Support from the Swedish grant agencies SSF OXIDE, VR, and KVA is
gratefully acknowledged.

\end{document}